**Natural mortality of *Trachurus novaezelandiae* and their size selection by purse seines off south-eastern Australia**

M. K. BROADHURST

*NSW Department of Primary Industries, Fisheries Conservation Technology Unit, PO Box 4321, Coffs Harbour, NSW 2450, Australia*

*Marine and Estuarine Ecology Unit, School of Biological Sciences, University of Queensland, Brisbane, QLD 4072, Australia*

M. KIENZLE

*Department of Agriculture and Fisheries, Ecosciences Precinct, 41 Boggo Road, Dutton Park, Brisbane, QLD 4072, Australia*

*University of Queensland, School of Agriculture and Food Sciences, St Lucia, QLD 4072, Australia.*

J. STEWART

*NSW Department of Primary Industries, Sydney Institute of Marine Science, Chowder Bay Road, Mosman, NSW 2088, Australia*

Correspondence: Matt Broadhurst, NSW Department of Primary Industries, Fisheries Conservation Technology Unit, PO Box 4321, Coffs Harbour, NSW 2450, Australia (email: matt.broadhurst@dpi.nsw.gov.au)

**Abstract** The natural mortality ($M$) and purse-seine catchability and selectivity were estimated for *Trachurus novaezelandiae*, Richardson, 1843 (yellowtail scad)—a small inshore pelagic species harvested off south-eastern Australia. Hazard functions were applied to two decades of data describing catches (mostly stable at a mean ±SE of 315 ± 14 t p.a.) and effort (declining from a maximum of 2289 to 642 boat days between 1999/00 and 2015/16) and inter-dispersed (over nine years) annual estimates of size-at-age (0+ to 18 years) to enable survival analysis. The data were best described by a model with eight parameters, including catchability (estimated at $< 0.1 \times 10^{-7}$ boat day$^{-1}$), $M$ (0.22 year$^{-1}$) and variable age-specific selection up to 6 years with a 50% retention among 5-year olds (larger than the estimated age at maturation). The low catchability implied minimal fishing mortality by the purse-seine fleet. Ongoing monitoring and applied gear-based studies are required to validate purse-seine catchability and selectivity, but the data nevertheless imply *T. novaezelandiae* could incur substantial additional fishing effort and, in doing, so alleviate pressure on other regional small pelagics.

**Introduction**

The genus *Trachurus* comprises 14 species of small- to medium-sized (~20–80 cm total length–TL) schooling fish that mostly have restricted distributions throughout tropical and temperate seas worldwide (Horn 1993; Stewart & Ferrell 2001; Yankov & Raykov 2012; Ward *et al*. 2016; Sexton *et al.* 2017). All 14 species are commercially or recreationally important, and most are harvested using various active (benthic and midwater trawls and purse seines; e.g. Horn 1993), and to a lesser extent, passive (traps and hook and line; e.g. Stewart & Ferrell 2001) gears.

Three *Trachurus* have distributions that encompass Australia (and also New Zealand), including *T. murphyi* Nichols, 1920 (Peruvian jack mackerel), *T. declivis* (Jenyns, 1841) (common jack mackerel) and *T. novaezelandiae* Richardson, 1843 (yellowtail scad) (Horn 1993; Stewart & Ferrell 2001). Their Australian distributions mostly encompass the southern half of the country, although *T. murphyi* also extends to the tropical north (Sexton *et al.* 2017). Like other *Trachurus*, the three species in Australia predominantly are targeted by purse seines and midwater trawls or taken as bycatch in benthic trawls targeting other teleosts or penaeids (Liggins *et al*. 1996). *Trachurus murphyi* and *T. declivis* collectively are managed in a national commercial fishery with a combined total allowable catch (TAC) of ~22 000 t p.a and are considered sustainable (AFMA, 2017). Owing to their economic importance, both species have had aspects of their life-histories assessed, but especially *T. declivis* which is by far the most abundant (e.g. Stevens & Hausfield 1982; Williams & Pullen 1993; Marshall *et al*. 1993; Sexton *et al*. 2017; Ward *et al.* 2017).

While *T. novaezelandiae* encompasses comparable distributions as its Australian conspecifics it has not incurred similar fishing effort, which might at least partially reflect its more estuarine and inshore distribution (i.e. accessibility). Australian landed catches of *T. novaezelandiae* mostly are restricted to New South Wales (NSW) and typically are between 400 and 600 t p.a.—up to 70% of which is harvested by small boats (5–15 m long) deploying purse seines with variable mesh sizes (stretched mesh openings) between 10 and 150 mm and headline lengths from 275 to 1000 m long (Stewart & Ferrell 2001). The recreational harvest of *T. novaezelandiae* is substantially smaller at ~15 to 60 t p.a. (Henry & Lyle 2003; West *et al.* 2015).

Despite their regional importance, relatively few primary literature studies have estimated key population parameters for *T. novaezelandiae* (but see Horn 1993; Stewart & Ferrell 2001; Neira *et al.* 2015) and consequently management remains contentious, primarily owing to concerns among some recreational fishers of poor stock sustainability and equitable harvesting (Neira 2009). Understanding basic life-history parameters and the vulnerability of a stock to capture by different fishing gears are precursors to the effective management of any fishery (Pauly 1980). It is especially important to obtain such parameters for schooling species like *Trachurus* that aggregate and are vulnerable at different stages to capture (e.g. during reproduction, feeding, etc.). Pivotal among population parameters is total mortality ($Z$), which typically is partitioned into that caused by natural ($M$) and fishing ($F$) causes and, ideally age-specific vulnerabilities (Wang 1999). Equally important, in terms of regulating fishing mortality, are gear selectivity assessments to control the exploitation rate (McLennan 1992).

Various analytical methods exist to estimate mortality and exploitation rates affecting aquatic organisms. Natural mortality can be simplistically derived via empirical relationships between life-history traits and environmental factors (Pauly, 1980), while $F$ often is estimated by subtracting $M$ from $Z$, with the latter obtained from catch curves (e.g. Stevens & Hausfeld 1982; Yankkova & Raykov 2012). For unexploited stocks, $Z$ also represents $M$, which was the approach used by Horn (1993) to postulate the only available estimates for *T. novaezelandiae* at 0.17–0.20 off New Zealand.

Where there are time-series estimates of size-at-age (i.e. derived from otoliths) several more complex and precise methods for calculating mortalities include linear regression on log-transformed age data (Millar 2014) and, more recently, hazard functions (to enable 'survival analysis'; Kienzle 2016; Broadhurst *et al.* 2017). Survival analysis has the utility of facilitating concurrent catchability and selectivity assessments, and can be done when only age samples and effort data are available (Kienzle 2016).

In NSW, *T. novaezelandiae* is characterized by historical data sets that meet the requirement for survival analyses. More specifically, otolith data have been collected by researchers across intermittent years for the past two decades, and there are accurate annual estimates of effort (boat days) and landed catches by purse seiners. Considering the above, our primary aim here was to quantify the mortality and purse-seine catchability and selectivity of *T. novaezelandiae* off New South Wales as a precursor to their ongoing effective management.

**Material and methods**

*Data used*

The data used to model the mortality and purse-seine catchability and selectivity of *T. novaezelandiae* were obtained from a long-term commercial-fishery monitoring program run by the NSW Department of Primary Industries between 1996 and 2016. The first data group was the number of aged individuals (0+–18 years) which comprised a total of 2347 otolith-based age estimates collected from representative samples of the commercial purse-seine fishery landings from nine of the 20 years (Table 1).

The second data group was annual yields for the two decades derived from reported commercial catch records partitioned for the various sectors (state and national commercial fisheries, including NSW purse seiners) and recreational harvests imputed from three creel surveys done during 1996/97, 2000/01 and 2013/14 (Henry & Lyle 2003; Steffe *et al*. 1996; West *et al*. 2015). As a result of potentially inadequate logbook reporting prior to 1997, the harvest from the purse-seine sector in 1996/97 (only) was estimated from the total catch in that year using the average of the total harvest reported by this sector between 1997/98 and 2015/16.

The third data group was annual fishing effort within the purse-seine sector (boat days) obtained from mandatory logbooks (Table 1), while the last data group comprised the average weight for each age group caught in each year calculated from individual weights collected during surveys of commercial landings. Several quantities subsequently were derived from these data following Broadhurst *et al.* (2017), including the: (1) proportion of fish in each age group; (2) estimated annual number of fish caught in each year; and (3) the estimated number of fish in each age group in each year.

*Models and mortality*

Initially, a stochastic population model was used to estimate the mortality of *T. novaezelandiae* by maximum likelihood using the matrix of age sample ($S$) based on a competing-risk model and first assuming two sources ($M$ and $F$) as described by Broadhurst *et al.* (2017). Various mortality models, represented by matrices $M$, $F$ and $Z$, where $z_{i,j} = m_{i,j} + f_{i,j}$, were fitted to determine which was best supported by the data. All models assumed the $M$ matrix was constant with all its elements equalled to $m$. An $F$ matrix was constructed according to the separability assumption as the product of: (1) gear selectivity ($G$); a matrix with rows representing the selectivity pattern at age; (2) fishing effort ($\tilde{E}$; a matrix with identical columns equal to $E$); and (3) a constant catchability matrix ($q$).

Based on the distribution of ages during the sampling years, all models assumed *T. novaezelandiae* aged 0–6 years could at least be partially selected by the purse seines. The remaining ages were assumed to be fully selected. Models were fitted to the data to estimate parametric characteristics for different hypotheses. The models included purse-seine selectivity at age constrained to vary between 0 and 100%. Different parametrisations were tested to compare non-parametric selectivity to the logistic function:

$$s(a) = \frac{1}{1+\exp(\alpha-\beta*a)} \qquad (1)$$

and a double-logistic function:

$$s(a) = \begin{cases} \frac{1}{1+\exp(\alpha-\beta*a)} & if\ a < \gamma \\ \frac{1}{1+\exp(\alpha-\beta*(\gamma-(a-\gamma)))} & if\ a \geq \gamma \end{cases} \qquad (2)$$

The first of these models estimated eight parameters including an age-by-age selectivity for age-groups up to 6 years old (i.e. catchability, $M$ and selectivity for age groups 0–1 to 5–6). The second model estimated catchability, $M$ and two parameters for the logistic gear selectivity. The third model estimated five parameters assuming double-logistic gear selectivity.

Minimization of the logarithm of the likelihood function, $L(\theta)$, was used to determine the best parameter estimates $\tilde{\theta}$ for the vector of parameters $\theta$. A region around the best likelihood estimates that contains the true values of the parameters with a certain probability ($W$) was defined according to Brandt (1998):

$$L(\theta) = L(\tilde{\theta}) + \frac{1}{2} \chi^2_W(n_f) \qquad (3)$$

where $L(\theta)$ is the value of the log-likelihood function for values $\theta$, $L(\tilde{\theta})$, is the minimum of the log-likelihood function and $\chi^2_W(n_f)$ is the quantile of the $\chi^2$-distribution for $n_f$ degrees of freedom and probability $W$. The 95% confidence intervals of parameters were calculated by profile likelihood using $W = 0.95$ and $n_f$ equal to the number of parameters in the model (Bolker 2008).

**Results**

*Temporal catch and effort*

The annual total catch of *T. novaezelandiae* remained fairly consistent at ~500 t during the two studied decades but did reach a low of 327 t in 2009–10 and a high of 650 t in 1997–98 (Table 1, Fig. 1). Throughout the two decades, ~50–70% of the total catch was taken by purse seines and with a broad temporal reduction in effort from a peak of 2289 boat days in 1999–00 to 642 boat days in 2015–16 (Fig. 1). These temporal trends manifested as an increase in nominal catch-per-unit-of-effort (CPUE) from around 200 kg boat day$^{-1}$ during the late 1990s to more than 400 kg boat day$^{-1}$ in the four most recent years.

*Model parameter estimates*

Of the assessed models, the first (model 1) incorporating age-specific purse-seine gear selectivity had the lowest Akaike information criterion (AIC), and was chosen as the best fit (Table 1). In this model, fleet selectivity was estimated to increase from 0% at age 0 to 100% at age 7, and with a ~50% retention among fish aged 5 years (Table 2, Figs 2 and 3). The model estimated $M$ at 0.22 year$^{-1}$ (or ~20% per year) with a 95% confidence interval ranging from 0.17 to 0.26 and catchability at $< 0.1 \times 10^{-7}$ boat day$^{-1}$ (0 to $463 \times 10^{-7}$) (Table 2, Fig. 3). The likelihood profile around the maximum likelihood estimate of catchability suggested that there is little information about this parameter in the dataset other than it must be very low (at $<463 \times 10^{-7}$ boat day$^{-1}$; Fig. 3) implying very low $F$. An increase in fishing power of 2 and 4% year$^{-1}$ was included into model 1, but this did not improve the fit.

**Discussion**

This study provides the first estimates of $M$, catchability and fishery selectivity for *T. novaezelandiae* off south-eastern Australia, contributing towards the limited relevant published work among both regional (Stewart & Ferrell 2001) and New Zealand (Horn 1993) conspecifics, and for the genera more broadly (e.g. Santic *et al.* 2002; Stevens & Hausfeld, 1982; Yankov & Raykov 2012; Yankov 2013). The parameters derived here represent an important first step for estimating exploitation and harvestable yields; albeit with some conditional assumptions.

Natural mortality was estimated to fall within a range of values (0.17–0.26 year$^{-1}$) encompassing those provided by Horn (1993) for conspecifics and the congeneric *T. declivis* (0.17–0.20 year$^{-1}$) off New Zealand, but substantially less than estimates for many *Trachurus* in other studies including *T. trachurus* ($M = 0.46$ year$^{-1}$, Santic *et al.* 2003); *T. declivis* ($M = 0.63$–0.70 year$^{-1}$; Stevens & Hausfeld 1982), and *T. mediterraneus* ($M = 0.86$ and 1.08 year$^{-1}$ Yankov & Raykov 2012; Yankov 2013). Some of these differences might be partially attributed to divergent modelling approaches, although Horn's (1993) similar intra-specific estimate was derived from catch curves (on an unexploited stock). Assuming validity among previous estimates, the lower $M$ for *T. novaezelandiae* than most of its congenerics probably reflects their divergent life histories. Specifically, *T. novaezelandiae* is a temperate inshore, coastal species, with adults inhabiting both estuarine and inshore waters generally <150 m (May & Maxwell 1986). Possibly, *T. novaezelandiae* evolved relatively protracted longevity and hence lower $M$ to ensure population persistence among an environment that is subjected to variable oceanographic and estuarine conditions.

Owing to the lack of any large industrial-type fishery for *T. novaezelandiae*, the estimated $M$ appears to represent most of their $Z$, and with negligible $F$. While *T. novaezelandiae* is exploited (often simply for bait use by recreational and commercial-line fishers) throughout its entire distribution, ~90% of the NSW commercial harvest is reported from only 5° of latitude (32–37°S) fished by the purse seiners (NSW DPI unpublished data). In addition, the majority of the purse-seine harvest is of larger fish destined for human consumption. As a consequence, the catchability was estimated to be extremely small: at ~$463 \times 10^{-7}$ boat day$^{-1}$. At this rate, it would take ~4600 purse-seine boat days to inflict a $F$ approaching the estimated $M$, or nearly up to a five-fold increase compared to effort expanded over the last decade.

It is also apparent that based on the estimated logistic selectivity for the purse seines (a model supported by other studies on similar gears; Stewart *et al*. 2005; Broadhurst *et al*. 2007) that any increase in effort would still be associated with some protection over juvenile $F$—assuming that most escaping fish survive (Broadhurst *et al*. 2006). More specifically, while there are no published estimates of age or size at maturity for *T. novaezelandiae*, Kaiola *et al*. (1996) suggested sizes at maturity for females and males at 20 and 22 cm fork length (FL), respectively. According to Stewart and Ferrell (2001), 2-year-old *T. novaezelandiae* attain a mean size of ~19 cm FL which, based on the chosen model here, implies that large proportions (e.g. 93% of all 2-year old fish) of maturing fish should avoid capture by purse seines.

While collective consideration of the parameters estimated here implies an underutilized fisheries resource, there are some limitations with respect to the analyses that warrant consideration. First, the size-at-age data were sporadic. Ideally, survival analysis supports long-term (5–7 consecutive years) of size-at-age data, although clearly based on meaningful model convergences here, gaps between years are feasible.

Second, purse-seine catchability and selectivity should be validated, which would be best done by applied work following empirical methods used among similar fishing gears (e.g. Stewart *et al.* 2005; Broadhurst *et al.* 2007).

Third, it is important to appreciate that various factors can affect $F$, including chronological variations (typically an increase) in fishing power (Robins *et al.* 1998; Bishop *et al.* 2008). Certainly, the observed increase in CPUE by the purse-seine fleet among recent years supports an enhanced capacity to target and catch *T. novaezelandiae*—especially considering the lack of evidence to suggest a concomitant increase in abundance. However, information concerning variable fishing power remains unavailable, and an annual increase in fishing power of 2 and 4% failed to improve model fit here.

Notwithstanding the caveats above, the data modelled here imply that the south-eastern Australian stock of *T. novaezelandiae* could withstand greater $F$ and, assuming appropriate markets, possibly might offset fishing pressure on either congenerics, or other small pelagic harvested by seines. Unlike some other mobile gears (e.g. demersal trawls) most seines, and especially purse seines, are associated with minimal bycatch, no habitats impacts and often relatively low energy requirements. Therefore, these gears fit within a broader model being promoted globally of 'low impact fuel efficient' (LIFE) methods as a mechanism to promote sustainable fisheries harvesting (Suuronen *et al.* 2012).

**Acknowledgements**

This study was funded by the NSW Department of Primary Industries and the Queensland Department of Agriculture and Fisheries.

**References**

AFMA (2017) AFMA small pelagic fishery. www.afma.gov.au/fisheries/small-pelagic-fishery/

Broadhurst M.K., Wooden M.E.L. & Millar R.B. (2007). Isolating selection mechanisms in beach seines. *Fisheries Research* **88**, 56–69.

Broadhurst M.K., Kienzle M. & Stewart J. (2017) Natural and fishing mortalities affecting eastern sea garfish, *Hyporhamphus australis* inferred from age-frequency data using hazard functions. Sub to *Fisheries Research.*

Bolker B.M. (2008) Ecological Models and Data in R. Princeton University Press.

Henry G.W. & Lyle J.M. (2003) The national recreational and indigenous fishing survey. Australian Government Department of Agriculture, Fisheries and Forestry, Canberra, Australia.

Horn P.L. (1993) Growth, age stucture, and productivity of jack mackerels (*Trachurus* spp.) in New Zealand waters. *New Zealand Journal of Marine and Freshwater Research* **27,** 145–155.

Kailola P.J., Williams M.J., Stewart P.C., Reichelt R.E., McNee A. & Grieve C (1993) Australian fisheries resources. Bureau of Resource Sciences, Australia & Fisheries Research and Development Corporation. 422 p.

Kienzle M. (2016). Hazard function models to estimate mortality rates affecting fish populations with application to the sea mullet (*Mugil cephalus*) fishery on the Queensland coast (Australia). *Journal of Agricultural, Environmental and Biological Statistics*, **21,** 76–91.

Liggins G.W., Kennelly S.J. & Broadhurst M.K. (1996) Observer-based survey of by-catch from prawn trawling in Botany Bay and Port Jackson, New South Wales. *Marine and Freshwater Research* **47,** 877–888.

Marshall J., Pullen G. & Jordan A. (1993) Reproductive biology and sexual maturity of female jack mackerel, *Trachurus declivis* (Jenyns), in eastern Tasmanian waters. *Australian Journal of Marine and Freshwater Research* **44,** 799–809.

May J.L. & Maxwell J.G.H. (1986) Trawl fish from temperate Waters of Australia CSIRO. Division of Fisheries Research, Tasmania. 492 pp.

MacLennan D.N. (1992) Fishing gear selectivity: an overview. *Fisheries Research* **13,** 201–204.

Millar R.B. (2014) A better estimator of mortality rate form age-frequency data. *Canadian Journal of Fisheries and Aquatic Sciences* **72,** 364–375.

Neira F.J. (2009) Provisional spawning biomass estimates of yellowtail scad (*Trachurus novaezelandiae*) off south-eastern Australia. New South Wales Department of Primary Industries Report, 32 pp.

Pauly D. (1980) On the interrelationships between natural mortality, growth parameters, and mean environmental temperature in 175 fish stocks. *Journal du Conseil International pour L'exploration de la Mer* **39,** 175–192.

R Core Team (2016) R: A Language and Environment for Statistical Computing. R Foundation for Statistical Computing, Vienna, Austria, 2016. URL http://www.R-project.org/.

Sexton S.C., Ward T.M. & Huveneers C. (2017) Characterising the spawning patterns of jack mackerel (*Trachurus declivis*) off eastern Australia to optimise future survey design. *Fisheries Research* **186,** 223–236.

Steffe A., Murphy S.J, Chapman D., Tarlington B.E., Gordon G.N.G., & Grinberg A. (1996) An assessment of the impact of offshore recreational fishing in New South Wales on the management of commercial fisheries. Project 94/053. Sydney, NSW Fisheries Research Institute: 139 pp.

Stevens J.D., Hausfeld H.F. (1982) Age determination and mortality estimates on an unexploited population of jack mackerel *Trachurus declivis* (Jenyns, 1841) from south-east Australia. CSIRO Marine Laboratories Report no. 148. 14 pp.

Stewart J., Ferrell D.J. (2001) Age, growth and commercial landings of yellowtail scad (*Trachurus novaezelandiae*) and blue mackerel (*Scomber australasicus*) off the coast of New South Wales, Australia. *New Zealand Journal of Marine and Freshwater Research* **35,** 541–551.

Stewart J., Walsh C., Reynolds D., Kendall B. & Gray C. (2004) Determining an optimal mesh size for use in the lampara net fishery for eastern sea garfish, *Hyporhampus australis*. *Fisheries Management and Ecology* **11,** 403–410.

Suuronen P., Chopin F., Glass C., Løkkeborg, S., Matsushita, Y., Queirolo, D. & Rihan D. (2012) Low impact and fuel efficient fishing: looking beyond the horizon. *Fisheries Research* **119,** 135–146.

Ward T.M., Burnell O.W., Ivey A., Sexton S.C., Carroll J., Keane J. & Lyle J.M. (2016) Spawning biomass of jack mackerel (*Trachurus declivis*) off eastern Australia: Critical knowledge for managing a controversial fishery. *Fisheries Research* **179,** 10–22.

Wang Y-G. (1999) A maximum-likelihood method for estimating natural mortality and catchability coefficient from catch-and-effort data. *Marine and Freshwater Research* **50,** 307–311.

West L.D., Stark K.E, Murphy J.J., Lyle J.M. & Doyle F.A. (2015) Survey of recreational fishing in New South Wales and the ACT, 2013/14. Fisheries Final Report Series.

Williams H. & Pullen G. (1993) Schooling behaviour of jack mackerel *Trachurus declivis* (Jenyns) observed in the Tasmanian purse seine fishery. *Australian Journal of Marine and Freshwater Research* **44,** 577–587.

Yankova M. & Raykov V. (2012) Growth, mortality and yield per recruit of horse mackerel (*Trachurus mediterraneus*) from the Bulgarian Black Sea waters. *Journal of Environmental Protection and Ecology* **13,** 1817–1823.

Yankova M. (2013) Population dynamics of horse mackerel (*Trachurus Mediterraneus ponticus*) in the Bulgarian Black Sea Coast. *Zoology*, Article ID 127287, 6 p

Table 1. Total harvested weights and purse-seine harvested weights and fishing (boat) days for *Trachurus novaezelandiae* and the numbers of individuals assessed for size at age (via otoliths) between 1996/97 and 2015/16 off New South Wales, Australia. –, no data collected

| Year | Total kg | Purse-seine kg | boat days | 0 | 1 | 2 | 3 | 4 | 5 | 6 | 7 | 8 | 9 | 10 | 11 | 12 | 13 | 14 | 15 | 16 | 17 | 18 |
|---|---|---|---|---|---|---|---|---|---|---|---|---|---|---|---|---|---|---|---|---|---|---|
| 1996/97 | 488 | 299 | 1209 | 0 | 5 | 326 | 110 | 42 | 22 | 64 | 22 | 12 | 2 | 5 | 2 | 0 | 0 | 0 | 0 | 0 | 0 | 0 |
| 1997/98 | 650 | 386 | 1397 | – | – | – | – | – | – | – | – | – | – | – | – | – | – | – | – | – | – | – |
| 1998/99 | 596 | 350 | 1781 | – | – | – | – | – | – | – | – | – | – | – | – | – | – | – | – | – | – | – |
| 1999/00 | 633 | 408 | 2289 | – | – | – | – | – | – | – | – | – | – | – | – | – | – | – | – | – | – | – |
| 2000/01 | 606 | 386 | 1341 | 0 | 0 | 0 | 6 | 15 | 18 | 36 | 21 | 26 | 37 | 48 | 25 | 38 | 18 | 11 | 14 | 5 | 2 | 8 |
| 2001/02 | 603 | 379 | 1165 | 0 | 8 | 42 | 58 | 43 | 38 | 24 | 39 | 42 | 24 | 13 | 8 | 7 | 4 | 2 | 2 | 1 | 0 | 0 |
| 2002/03 | 504 | 294 | 1094 | – | – | – | – | – | – | – | – | – | – | – | – | – | – | – | – | – | – | – |
| 2003/04 | 498 | 282 | 947 | 0 | 0 | 10 | 19 | 49 | 15 | 26 | 19 | 16 | 56 | 28 | 10 | 4 | 16 | 3 | 2 | 1 | 0 | 0 |
| 2004/05 | 463 | 263 | 915 | – | – | – | – | – | – | – | – | – | – | – | – | – | – | – | – | – | – | – |
| 2005/06 | 466 | 281 | 1012 | 0 | 37 | 1 | 17 | 36 | 35 | 16 | 18 | 6 | 6 | 5 | 9 | 11 | 0 | 0 | 0 | 4 | 0 | 0 |
| 2006/07 | 476 | 299 | 935 | 0 | 0 | 6 | 5 | 16 | 49 | 38 | 18 | 13 | 11 | 17 | 3 | 4 | 4 | 0 | 0 | 1 | 0 | 0 |
| 2007/08 | 446 | 266 | 801 | 5 | 9 | 1 | 9 | 6 | 20 | 54 | 33 | 27 | 9 | 15 | 2 | 3 | 12 | 5 | 0 | 4 | 0 | 0 |
| 2008/09 | 483 | 281 | 953 | – | – | – | – | – | – | – | – | – | – | – | – | – | – | – | – | – | – | – |
| 2009/10 | 327 | 227 | 714 | – | – | – | – | – | – | – | – | – | – | – | – | – | – | – | – | – | – | – |
| 2010/11 | 504 | 299 | 833 | – | – | – | – | – | – | – | – | – | – | – | – | – | – | – | – | – | – | – |
| 2011//12 | 433 | 219 | 764 | 0 | 0 | 5 | 21 | 34 | 36 | 8 | 1 | 2 | 3 | 2 | 3 | 2 | 1 | 0 | 0 | 0 | 0 | 0 |
| 2012/13 | 570 | 367 | 880 | 0 | 0 | 1 | 3 | 16 | 24 | 5 | 1 | 2 | 3 | 2 | 2 | 1 | 0 | 0 | 0 | 0 | 0 | 0 |
| 2013/14 | 592 | 427 | 967 | – | – | – | – | – | – | – | – | – | – | – | – | – | – | – | – | – | – | – |
| 2014/15 | 513 | 347 | 773 | – | – | – | – | – | – | – | – | – | – | – | – | – | – | – | – | – | – | – |
| 2015/16 | 412 | 244 | 642 | – | – | – | – | – | – | – | – | – | – | – | – | – | – | – | – | – | – | – |

Table 2. Results of fitted models (ranked by increasing value of AIC) estimating catchability, natural mortality ($M$) and selectivity by age class for *Trachurus novaezelandiae* off New South Wales Australia between 1996 and 2015. Parameter estimates for the chosen model are provided in parenthesis with 95% likelihood profile confidence intervals. Log, log-likelihood, Par, number of parameters, $M$, natural mortality, *s*, selectivity by age class, $\alpha$, $\beta$, $\gamma$

| Model | Par | AIC | Catchability | $M$ | *s* age 1 | *s* age 2 | *s* age 3 | *s* age 4 | *s* age 5 | *s* age 6 | $\alpha$ | $\beta$ | $\gamma$ |
|---|---|---|---|---|---|---|---|---|---|---|---|---|---|
| 1 | 8 | 10494.9 | $<0.1 \times 10^{-7}$ ($0; 463 \times 10^{-7}$) | 0.22 (0.17; 0.26) | 0 (0:0) | 0.07 (0.04; 0.13) | 0.35 (0.23; 0.51) | 0.36 (0.24; 0.51) | 0.63 (0.45; 0.84) | 0.98 (0.73; 1.00) | NA | NA | NA |
| 3 | 5 | 10654.8 | $<0.01 \times 10^{-7}$ ($0; 900 \times 10^{-7}$) | 0.17 (0.05; 0.24) | NA | NA | NA | NA | NA | NA | 3.43 (2.73; 4.54) | 1.38 (0.96; 2.50) | 9.6 (8.5; 18.0) |
| 2 | 4 | 10665.1 | $\leq 0.01 \times 10^{-7}$ ($0; 396 \times 10^{-7}$) | 0.19 (0.15; 0.23) | NA | NA | NA | NA | NA | NA | 3.54 (3.11; 4.07) | 1.31 (1.03; 1.94) | NA |

**Captions to figures**

Fig. 1. Annual total and purse-seined catches (t) and purse-seine fishing effort (boat days) for *Trachurus novaezelandiae* off New South Wales between 1996 and 2016.

Fig. 2. Estimated purse-seine selectivity by age for the three assessed models.

Fig. 3. Profile likelihoods for each of the eight parameters in the chosen model that best fit the data for *Trachurus novaezelandiae* off New South Wales between 1996 and 2016, including (a) catchability, (b) natural mortality, and selectivity of age-groups (c) 1, (d) 2, (e) 3, (f) 4, (g) 5 and (h) 6. The horizontal lines show the minimum negative log-likelihood (plus $\chi^2_{0.95}$ for 4 degrees of freedom) providing the boundaries of the 95% confidence internal for each parameter.

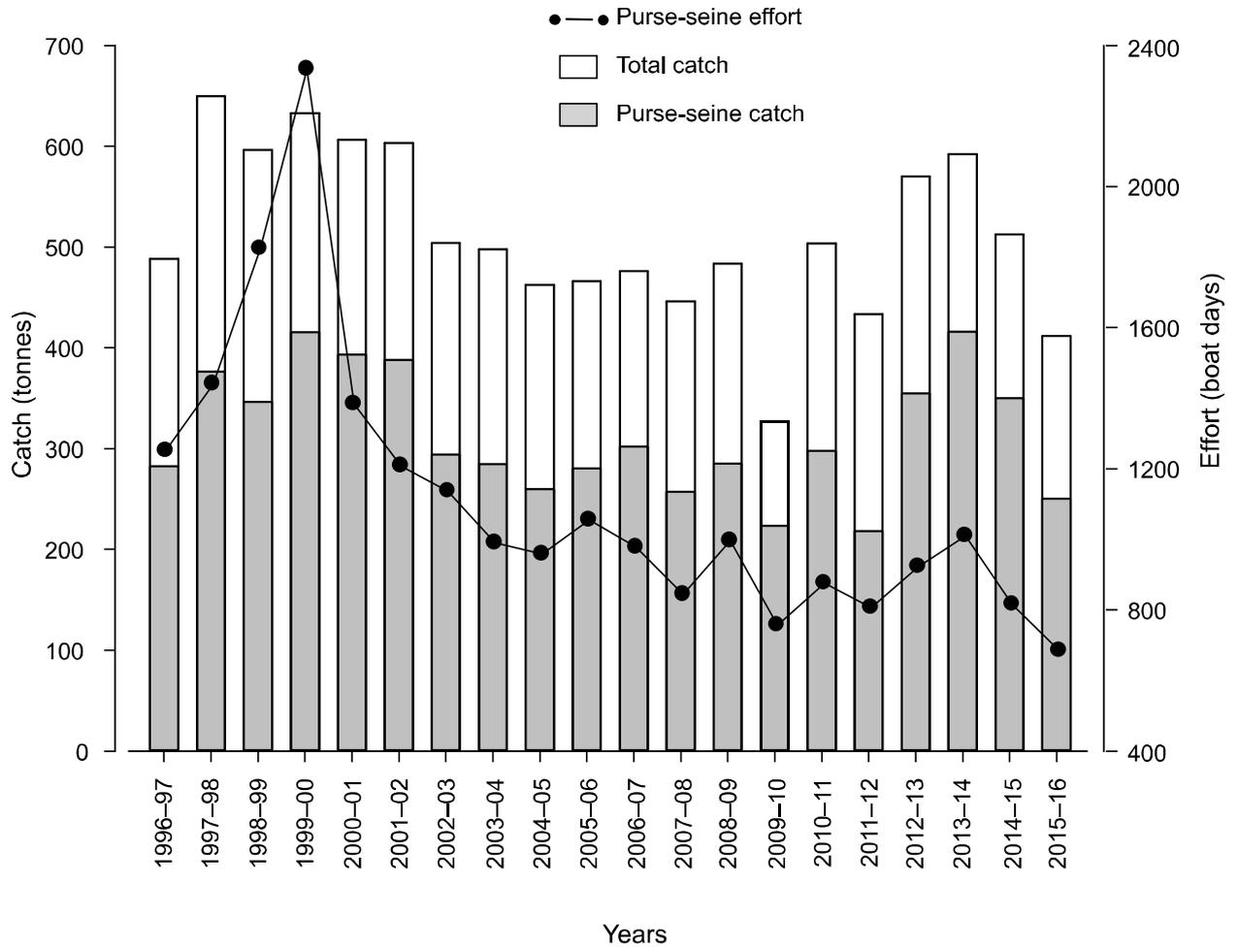

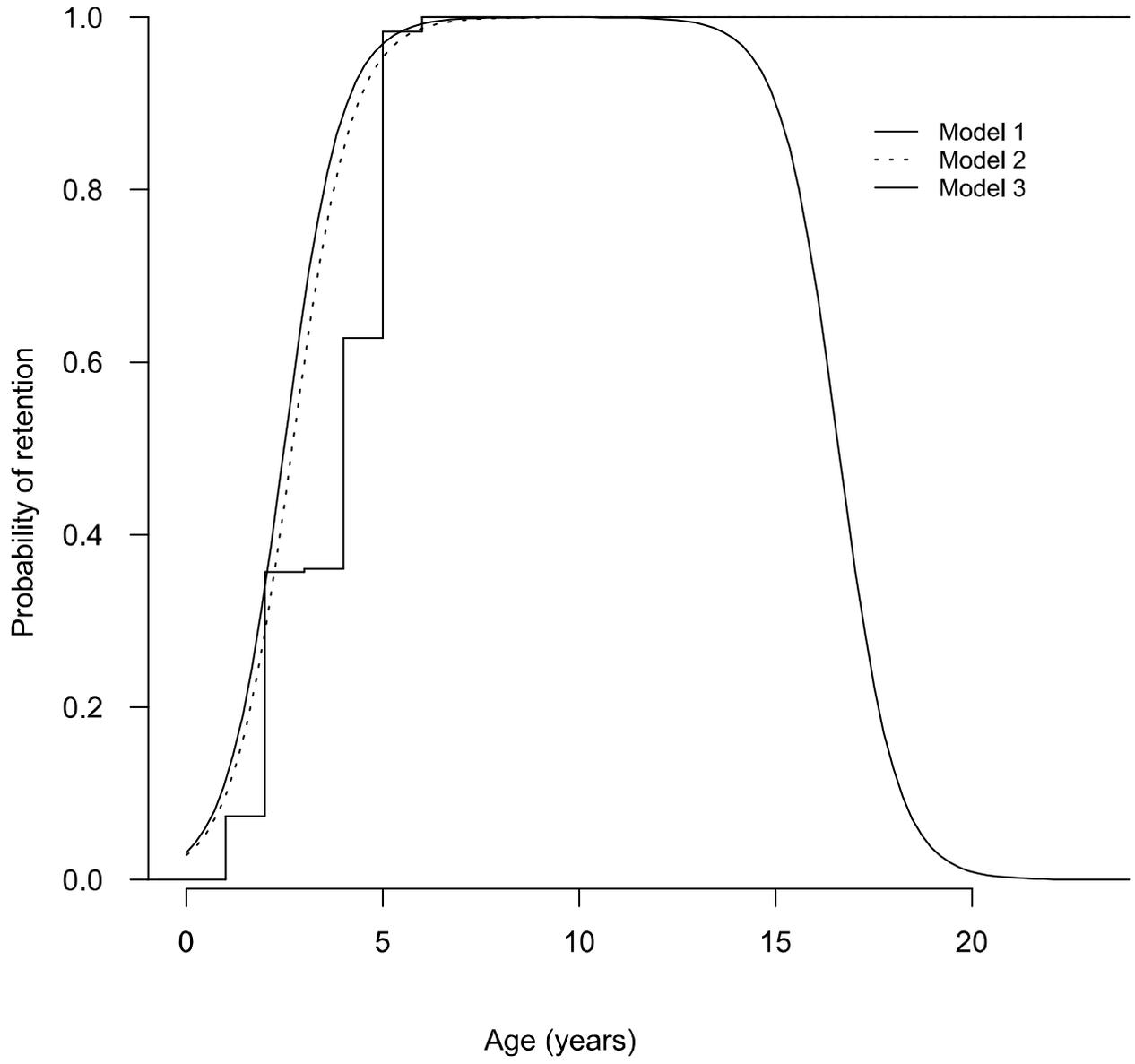

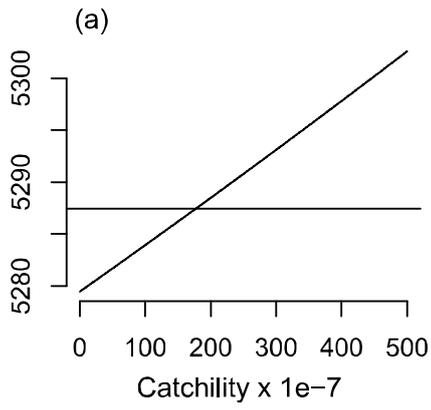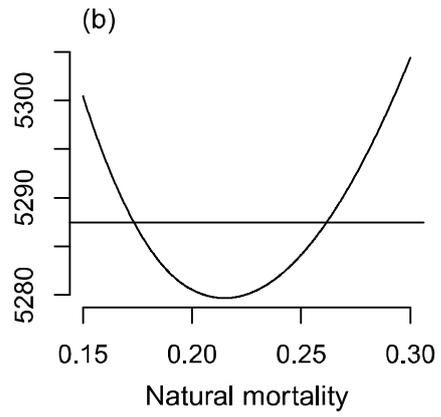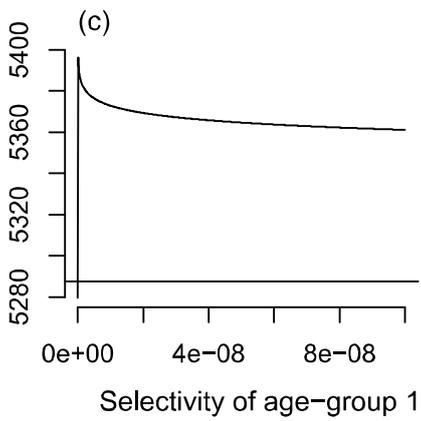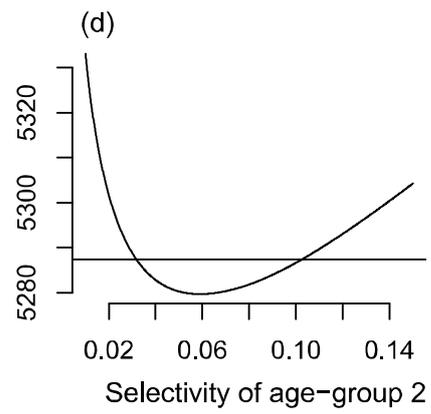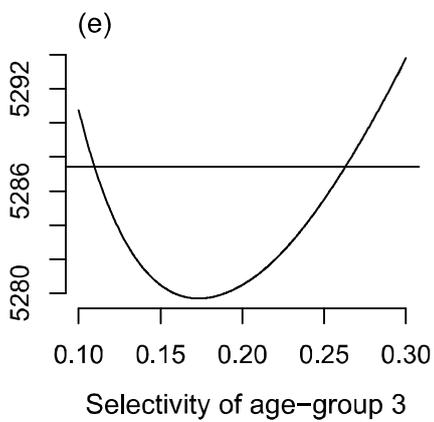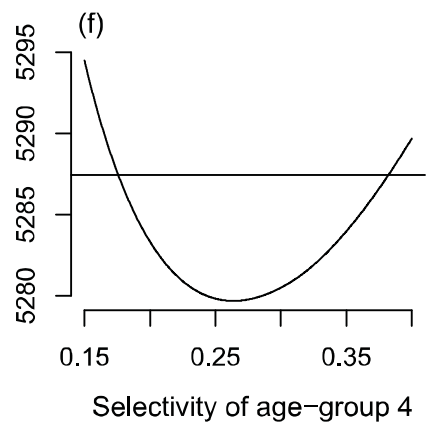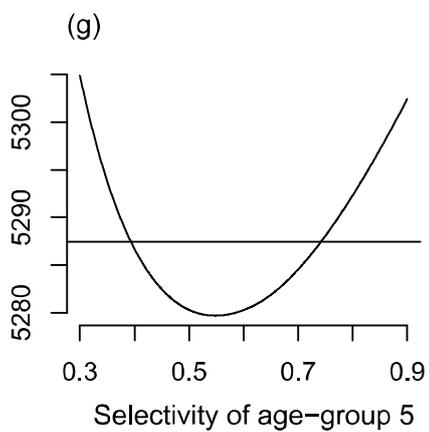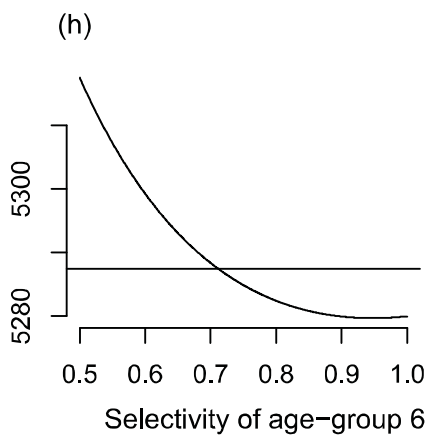